\documentclass[onecolumn,aps,pra,nofootinbib,superscriptaddress,tightenlines,notitlepage,longbibliography,12pt]{revtex4-1}

\usepackage{graphicx}
\usepackage{amsmath}
\usepackage{amsthm}
\usepackage{amsfonts}
\usepackage{color}
\usepackage{subfigure}
\usepackage{booktabs}
\usepackage{algpseudocode}

\usepackage[ breaklinks,
			 colorlinks = true,
             linkcolor = blue,
             urlcolor  = blue,
             citecolor = red,
             anchorcolor = blue,
]{hyperref}

\newcommand{\SHA}{\textrm{SHA}\ensuremath{256}}
\newcommand{\marketcap}{150 }

\begin{document}

\title{Quantum attacks on Bitcoin, and how to protect against them}

%%%%%

\author{Divesh Aggarwal}
\affiliation{National University of Singapore, Singapore}
\affiliation{Centre for Quantum Technologies, National University of Singapore, Singapore}

\author{Gavin K. Brennen}
\affiliation{Macquarie University, Sydney, Australia}

\author{Troy Lee}
\affiliation{SPMS, Nanyang Technological University, Singapore}
\affiliation{Centre for Quantum Technologies, National University of Singapore, Singapore}

\author{Miklos Santha}
\affiliation{CNRS, IRIF, Universit\'{e} Paris Diderot, France}
\affiliation{Centre for Quantum Technologies, National University of Singapore, Singapore}

\author{Marco Tomamichel}
\affiliation{University of Technology Sydney, Australia}

\begin{abstract}
The key cryptographic protocols used to secure the internet and financial transactions of today are all susceptible 
to attack by the development of a sufficiently large quantum computer.  One particular area at risk are cryptocurrencies, 
a market currently worth over \marketcap billion USD.  We investigate the risk of Bitcoin, and other cryptocurrencies, to attacks by quantum computers.  We find that the proof-of-work used by Bitcoin is relatively resistant 
to substantial speedup by quantum computers in the next 10 years, mainly because specialized ASIC miners are 
extremely fast compared to the estimated clock speed of near-term quantum computers.  On the other hand, the elliptic 
curve signature scheme used by Bitcoin is much more at risk, and could be completely broken by a 
quantum computer as early as 2027, by the most optimistic estimates.

We analyze an alternative proof-of-work called Momentum, based on finding collisions in a hash function, that is even 
more resistant to speedup by a quantum computer.  We also review the available post-quantum signature schemes to see which one 
would best meet the security and efficiency requirements of blockchain applications.    
\end{abstract}

\maketitle

\section{Introduction}
Bitcoin is a decentralized digital currency secured by cryptography.  Since its development by Satoshi Nakamato in 2008 \cite{Nak09},
Bitcoin has proven to be a remarkably successful and secure system and has inspired the development of hundreds of other 
cryptocurrencies and blockchain technologies in a market currently worth over \marketcap billion USD.

The security of Bitcoin derives from several different features of its protocol.  The first is the proof-of-work that is required to write 
transactions to the Bitcoin digital ledger.  The work required to do this safeguards against malicious parties who possess less than 
50\% of the computational power of the network from creating an alternative 
history of transactions.  The second is the cryptographic signature that is used to authorize transactions.  Bitcoin currently uses a signature 
scheme based on elliptic curves.

The coming development of quantum computers poses a serious threat to almost all of the cryptography currently used to secure the 
internet and financial transactions, and also to Bitcoin.  In this paper, we comprehensively analyse the vulnerability of Bitcoin to quantum 
attacks.  We find that the proof-of-work used by Bitcoin is relatively resistant 
to substantial speedup by quantum computers in the next 10 years, mainly because specialized ASIC miners are 
extremely fast compared to the estimated clock speed of near-term quantum computers.  This means that transactions, once on the 
blockchain, would still be relatively protected even in the presence of a quantum computer.  

The elliptic curve signature scheme used by Bitcoin is well-known to be broken by Shor's algorithm \cite{shor99} for computing discrete 
logarithms.  We analyse exactly \emph{how long} it might take to derive the secret key from a published public key on a future quantum 
computer.  This is critical in the context of Bitcoin as the main window for this attack is from the time a transaction is broadcast until the 
transaction is processed into a block on the blockchain with several blocks after it.  By our most optimistic estimates, as early as 2027 
a quantum computer could exist that can break the elliptic curve signature scheme in less than 10 minutes, the block time used in Bitcoin.  

We also suggest some countermeasures that can be taken to secure Bitcoin against quantum attacks.  We analyse an alternative 
proof-of-work scheme called Momentum \cite{Larimer14}, based on finding collisions in a hash function, and show that it admits even 
less of a quantum speedup than the proof-of-work used by Bitcoin.  We also review alternative signature schemes that are believed 
to be quantum safe.

\section{Blockchain basics}

In this section we give a basic overview of how bitcoin works, so that we can refer to specific parts of the protocol when 
we describe possible quantum attacks.  We will keep this discussion at an abstract level, as many of the principles apply 
equally to other cryptocurrencies with the same basic structure as bitcoin.  

All Bitcoin transactions are stored in a public ledger called the \emph{blockchain}.  Individual transactions are bundled 
into blocks, and all transactions in a block are considered to have occurred at the same time.  A time ordering is placed 
on these transactions by placing them in a chain.  Each block in the chain (except the very first, or \emph{genesis} block) 
has a pointer to the block before it in the form of the hash of the previous block's header.  

Blocks are added to the chain by \emph{miners}.  Miners can bundle unprocessed transactions into a block and 
add them to the chain by doing a \emph{proof-of-work} (PoW).  Bitcoin, and many other coins, use a PoW 
developed by Adam Back called Hashcash \cite{back02}.   The hashcash PoW is to find a well-formed \emph{block header}
such that $h(\text{header}) \le t$, where $h$ is a cryptographically secure hash function and $\text{header}$ is the block 
header.  A well-formed header contains summary information of a block such as a hash 
derived from transactions in the block \footnote{Specifically the root of a Merkle tree of hashes of the transactions}, a hash 
of the previous block header, a time stamp, as well as a so-called \emph{nonce}, a 32-bit register that can be freely 
chosen.  An illustration of a block can be found in Table~\ref{block_figure}.  The parameter $t$ is a target value that can be changed to 
adjust the difficulty of the PoW.  In Bitcoin, this parameter is dynamically adjusted every 2016 blocks such that the network 
takes about 10 minutes on average to solve the PoW.  

\begin{table}
\centering
\begin{tabular}{| l l |}
\hline
Version & 0x20000012 \\
Previous block header hash & $00\ldots0$dfff7669865430b\ldots \\
Merkle Root & 730d68233e25bec2\ldots \\ 
Timestamp & 2017-08-07 02:12:18 \\
Difficulty & 860,221,984,436.22 \\
Nonce & 941660394 \\
\hline
\multicolumn{2}{|c|}{Transaction 1} \\
\multicolumn{2}{|c|}{Transaction 2} \\
\multicolumn{2}{|c|}{\vdots} \\
\hline
\end{tabular}
\caption{Illustration of a block.  The data in the top constitutes the block header.}
\label{block_figure}
\end{table}

In Bitcoin the hash function chosen for the proof of work is two sequential applications of the 
$\SHA{}: \{0,1\}^* \rightarrow \{0,1\}^{256}$ hash function, i.e.\ $h(\cdot) = \SHA(\SHA(\cdot))$.  As the size of the range of $h$ is then $2^{256}$, the 
expected number of hashes that need to be tried to accomplish the hashcash proof of work with parameter $t$ is 
$2^{256}/t$.  Rather than $t$, the Bitcoin proof-of-work is usually specified in terms of the \emph{difficulty} $D$ where 
$D = 2^{224}/t$.  This is the expected number of hashes needed to complete the proof of work divided by $2^{32}$, the 
number of available nonces.  In other words, the difficulty is the expected number of variations of transactions and time 
stamps that need to be tried when hashing block headers, when for each fixing of the transactions and time stamp 
all nonces are tried.

Miners can bundle unprocessed transactions into a block however 
they like, and are awarded a number of bitcoins for succeeding in the PoW task.  The ``generation'' transaction paying the 
mining reward is also a transaction included in the block, ensuring that different miners will be searching over disjoint 
block headers for a good hash pre-image.

Once a miner finds a header satsifying $h(\text{header}) \le t$, they announce this to the network and the block is added to 
the chain.  Note that it is easy to verify that a claimed $\text{header}$ satisfies the PoW condition\,---\,it simply requires one 
evaluation of the hash function.  

The purpose of the PoW is so that one party cannot unilaterally manipulate the blockchain in order to, for 
example, double spend.  It is possible for the blockchain to fork, but at any one time the protocol dictates that miners 
should work on the fork that is currently the longest.  Once a block has $k$ many blocks following it in the longest chain, 
a party who wants to create a longest chain not including this block would have to win a PoW race 
starting $k$ blocks behind.  If the party controls much less than half of the computing power of the network, this becomes 
very unlikely as $k$ grows.  In Bitcoin, a transaction is usually considered safe once it has $6$ blocks following it.

The first question we will look at in Section~\ref{miningattack} is what advantage a quantum computer would have in 
performing the hashcash PoW, and if it could unilaterally ``come from behind'' to manipulate the blockchain.

The second aspect of Bitcoin that is important for us is the form that transactions take.  When Bob wants to send bitcoin 
to Alice, Alice first creates (an ideally fresh) private-public key pair.  The public key is hashed to create an 
\emph{address}.  This address is what Alice provides to Bob as the destination to send the bitcoin.  Bitcoin uses the 
hash of the public key as the address instead of the public key\footnote{In early versions of the Bitcoin protocol the 
public key could be used as an address.} not for security reasons but simply to save space.  As we 
see later, this design choice does have an impact on the \emph{quantum} security.

To send bitcoin to Alice, Bob must also point to transactions on the blockchain where bitcoin was sent to addresses that he 
controls.  The sum of bitcoin received to these referenced transactions must add up to at least the amount of bitcoin Bob 
wishes to send to Alice.  Bob proves that he owns these addresses by stating the public key corresponding to each 
address and using his private key corresponding to this address to sign the message saying he is giving these bitcoins to 
Alice.

\section{Quantum attacks on bitcoin}
\label{sec:attacks}

\subsection{Attacks on the Bitcoin proof-of-work}
\label{miningattack}
In this section, we investigate the advantage a quantum computer would have in performing the hashcash PoW used by 
Bitcoin.  Our findings can be summarized as follows: Using Grover search \cite{grover96}, a quantum computer can 
perform the hashcash PoW by performing quadratically fewer hashes than is needed by a classical computer.  However, the extreme speed of current specialized ASIC hardware for 
performing the hashcash PoW, coupled with much slower projected gate speeds for current quantum architectures, 
essentially negates this quadratic speedup, at the current difficulty level, giving quantum computers no advantage.  Future 
improvements to quantum technology allowing gate speeds up to 100GHz could allow quantum computers to solve the 
PoW about 100 times faster than current technology.  However, such a development is unlikely in the next decade, 
at which point classical hardware may be much faster, and quantum technology might be so widespread that no single 
quantum enabled agent could dominate the PoW problem.

We now go over these results in detail.  Recall that the Bitcoin PoW task is to find a valid block header such that 
$h(\text{header}) \le t$, where $h(\cdot) = \SHA(\SHA(\cdot))$. The security of the blockchain depends on no agent being able to solve the 
PoW task first with probability greater than 
$50\%$.  We will investigate the amount of classical computing power that would be needed to match 
one quantum computer in performing this task.

We will work in the random oracle model \cite{BR93}, and in particular assume that 
$\Pr[h(\text{header}) \le t] = t/2^{256}$ where the probability is taken 
uniformly over all well-formed block headers that can be created with transactions available in the pool at any given time (such well-formed block headers can be found by varying the nonce, the transactions included in the block as well as the least significant bits of the timestamp of the header). On a classical computer, the expected number of block headers and nonces which need to be hashed in order to find one whose hash value is at most $t$ is $D\times 2^{32}$ where $D$ is the hashing difficulty defined by $D=2^{224}/t$ \footnote{According to \href{https://blockchain.info/charts/difficulty?timespan=30days}{blockchain.info}, on August 8, 2017, the hashing difficulty was $D=860 \cdot 10^9$ and target was $t=2^{184.4}$}.

For quantum computers in the random oracle model we can restrict our attention to the generic quantum approach to solving the PoW task using Grover's algorithm~\cite{grover96}.  By Grover's algorithm, searching a database of $N$ items for a marked item can be done with $O(\sqrt{N})$ many queries to the database (whereas any classical computer would require $\Omega(N)$ queries to complete the same task).

Let $N=2^{256}$ be the size of the range of $h$ for the following.  By our assumptions, with probability at least $0.9999$ a random set of $10\cdot N/t$ many block headers will contain at least one element whose hash is at most $t$.  We can fix some deterministic 
function $g$ mapping $S=\{0,1\}^{\lceil{\log(10\cdot N/t)}\rceil}$ to distinct well-formed block headers.  We also define 
a function $f$ which determines if a block header is ``good'' or not
\[
f(x)=\left\{\begin{array}{c}0\quad{\rm if}\quad { h(g(x)) > {\it t}}\\ 1\quad{\rm if}\quad {h(g(x)) \le {\it t}}\end{array}\right..
\]
A quantum computer can compute $f$ on a superposition of inputs, i.e.\ perform the mapping
\[
\displaystyle{\sum_{x \in S} \alpha_x |x\rangle\rightarrow \sum_{x \in S}}(-1)^{f(x)} \alpha_x |x\rangle.
\]
Each application of this operation is termed an \emph{oracle call}.  Using
Grover's algorithm a quantum algorithm can search through $S$ to find a good block header by computing 
$\#\mathcal{O}=\frac{\pi}{4}\sqrt{10\cdot N/t}=\pi 2^{14}\sqrt{10\cdot D}$ 
oracle calls.  The Grover algorithm can be adapted to run with this scaling even if the number of solutions is not known beforehand, and even if no solutions exist \cite{Boyer:1998fk}. 

While the number of oracle calls determines the number of hashes that need to be performed, additional overhead will be incurred to compute 
each hash, construct the appropriate block header, and to do quantum error correction.  We now analyze these factors to 
determine a more realistic estimate of the running time in two ways. First, we estimate the running time based on a well studied model for universal quantum computers with error correction.

On a classical computer, a hash function such as \SHA{} uses basic boolean gate operations, whereas on a quantum computer, these elementary boolean gates are translated into reversible logical quantum gates which introduces some overhead. There are a total of 64 rounds of hashing in the \SHA{} protocol and each round can be done using an optimized circuit with $683$ Toffoli quantum gates \cite{Parent2015ReversibleCC}. Most quantum error correction codes use $T$ gates rather than Toffoli gates as the representative time consuming gate, and a careful analysis of the cost to perform the SHA256 function call as well as the inversion about the mean used in the Grover algorithm finds a total $T$ gate count of $474168$ for one oracle call \cite{Suchara:EECS-2013-119}. In that circuit decomposition, the $T$ gates can only be parallelized by roughly a factor of three.

There is additional overhead needed by quantum computers to perform error correction. In deciding on a good quantum error correction code there are a variety of tradeoffs to consider: tolerance to a particular physical error model, versatility for subroutines, number of qubits used, logical gate complexity, and the amount of classical processing of error syndromes and feedback. Adopting the surface code, which has advantages of a relatively high fault tolerance error threshold and local syndrome measurements, we can adapt the analysis in Ref. \cite{Suchara:EECS-2013-119} to estimate the total run time of the quantum algorithm. 
The time needed to run the Grover algorithm and successfully mine a block is
\begin{align*}
\tau &= \#\mathcal{O} \times \#G/s = \pi 2^{14}\sqrt{10\cdot D}\times \#G/s,
\end{align*}
where $\# G$ is the number of cycles needed for one oracle call, and $s$ is the quantum computer clock speed.
Using a surface code, where the dominant time cost is in distilling magic states to implement $T$ gates, one finds
\[
\# G= 297784 \times c_{\tau}(D,p_g),
\]
where the first factor includes the logical $T$ gate depth for calling the SHA256 function twice as required by bitcoin PoW, and twice again to make the circuit reversible, as well as the inversion about the mean. The second factor, $c_{\tau}$, is the overhead factor in time needed for quantum error correction. It counts the number of clock cycles per logical $T$ gate and is a function of difficulty and the physical gate error rate $p_g$. For a fixed gate error rate, the overhead factor $c_{\tau}$ is bounded above by the cost to invert a $256$ bit hash (maximum difficulty).
  \begin{figure}[t]
   \begin{center}
     \subfigure[]{
       \includegraphics[width=0.48\textwidth]{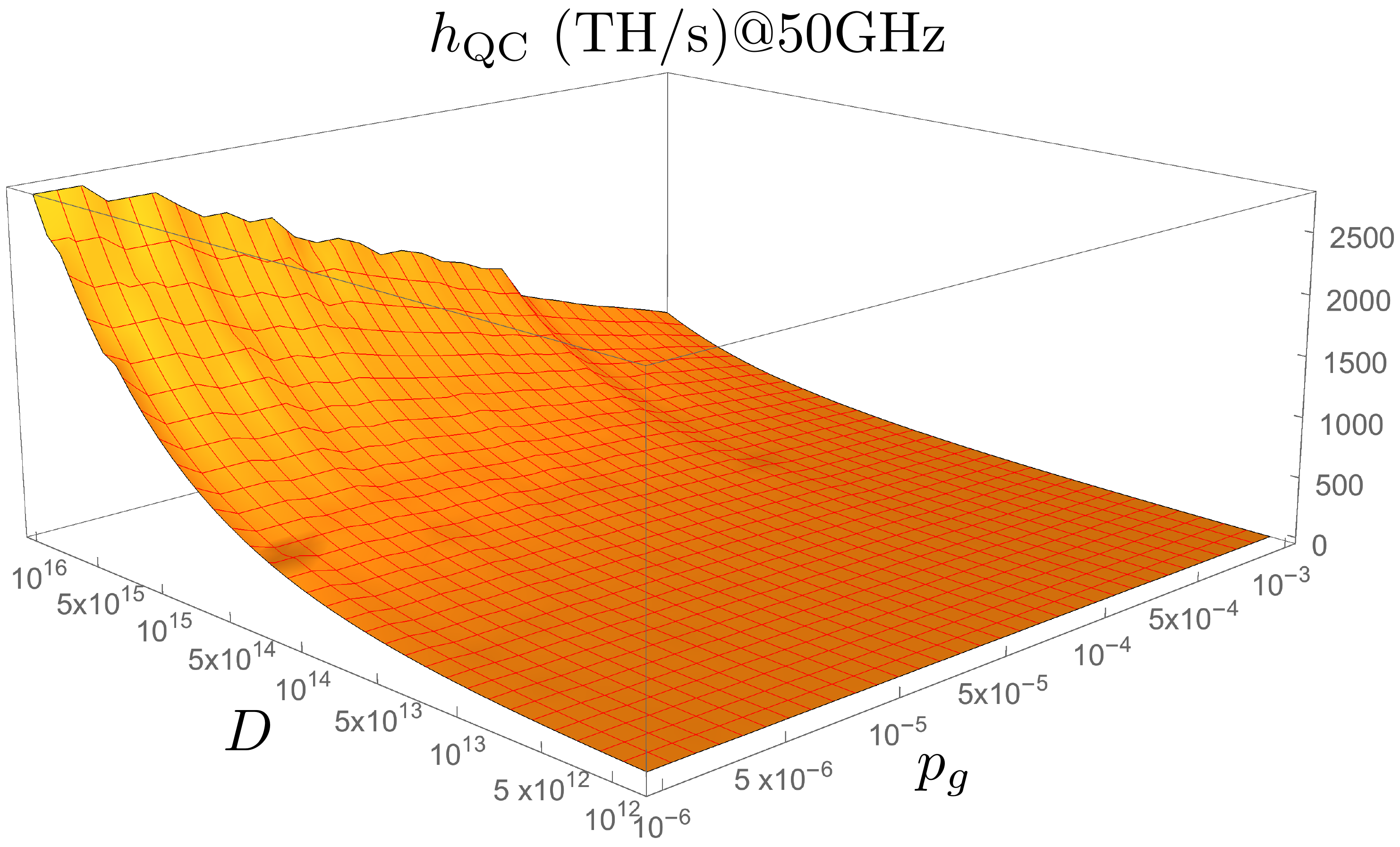}
       %\label{fig:numQ}
       }
     \subfigure[]{
       \includegraphics[width=0.48\textwidth]{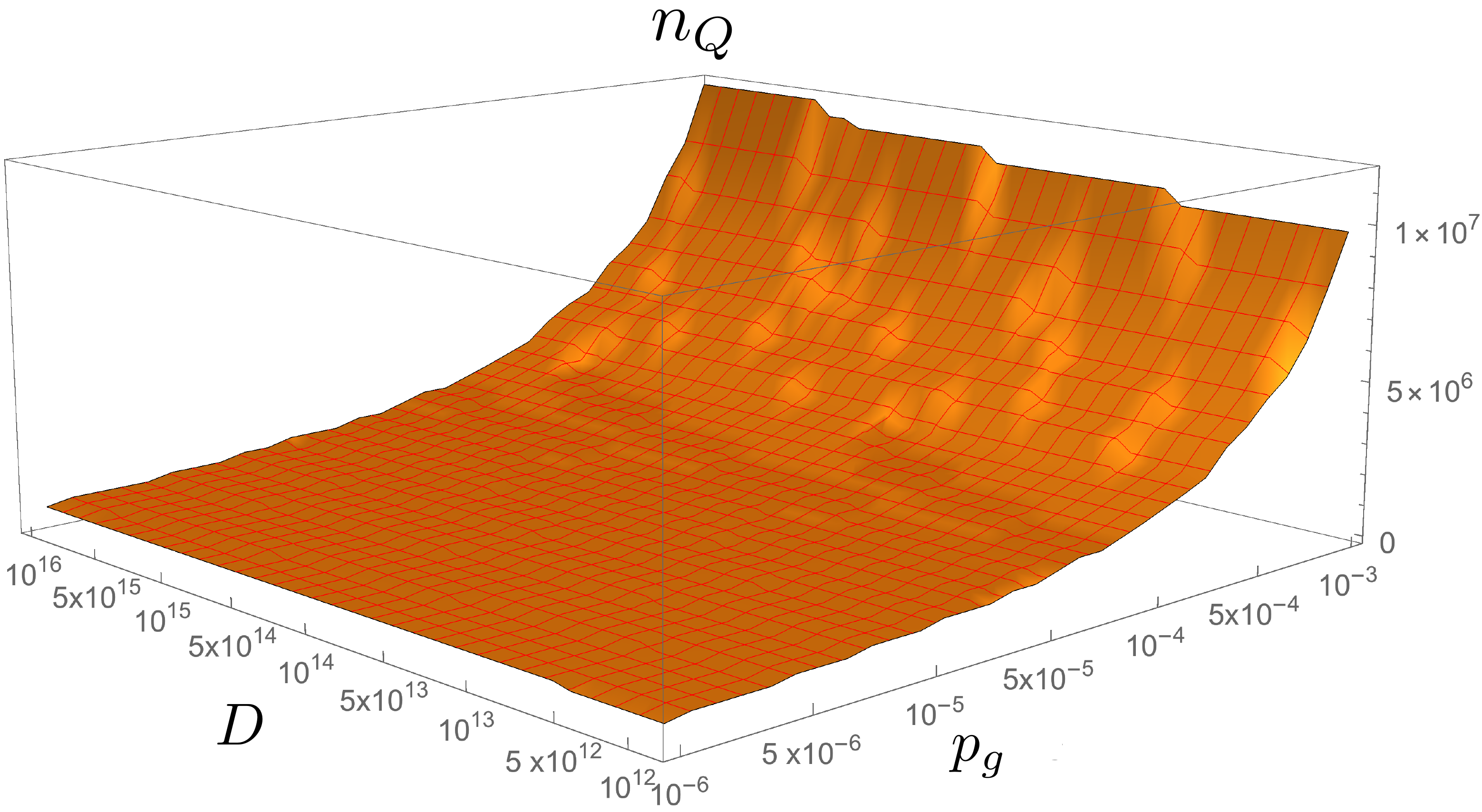}
       %\label{fig:hashrateQC}
       }
   \end{center}
   \caption{Performance of a single quantum computer for blockchain attacks as a function of physical gate error rate $p_g$, which is an internal machine specification, and mining Difficulty $D$, which is set by the blockchain protocol. (a) Effective hash rate $h_{\rm QC}$ for a quantum computer operating at $50$GHz clock speed which is at the optimistic limit of forseeable clock speeds. The hash rate increases as the square root of difficulty (note the log scale). For $d$ quantum computers acting in parallel the effective hash rate increases by a factor of $2.56\times \sqrt{d}$.(b) Number of physical qubits $n_{\rm Q}$ used by the quantum computer.}
   \label{fig:mineattack}
 \end{figure}
 
Because the quantum algorithm runs the hashing in superposition, there is no direct translation of quantum computing power into a hashing rate. However, we can define an effective hash rate, $h_{\rm QC}$, as the expected number of calls on a classical machine divided by the expected time to find a solution on the quantum computer, viz.
\[
h_{\rm QC}\equiv \frac{N/t}{\tau}=\frac{0.28\times s\sqrt{D}}{c_{\tau}(D,p_g)}.
\]
Because the time overhead is bounded, asymptotically the effective hashing rate improves as the square root of the difficulty, reflecting the quadratic advantage obtainable from quantum processors. 

The Grover algorithm can be parallelized over $d$ quantum processors. In the optimal strategy, 
each processor is dedicated to search over the entire space of potential solutions, and the expected number of oracle calls needed to find a solution is $\#\mathcal{O}=0.39\times \#\mathcal{O}/\sqrt{d}$ \cite{Gingrich:2000by}. This implies an expected time to find a solution is
\[
\tau_{\rm \|}=0.39\times \tau/\sqrt{d},
\]
and the effective hash rate using $d$ quantum processors in parallel is
\[
h_{\rm QC, \|}=2.56\times h_{\rm QC}\sqrt{d}.
\]
\begin{figure}[t]
  \includegraphics[width=0.75\textwidth]{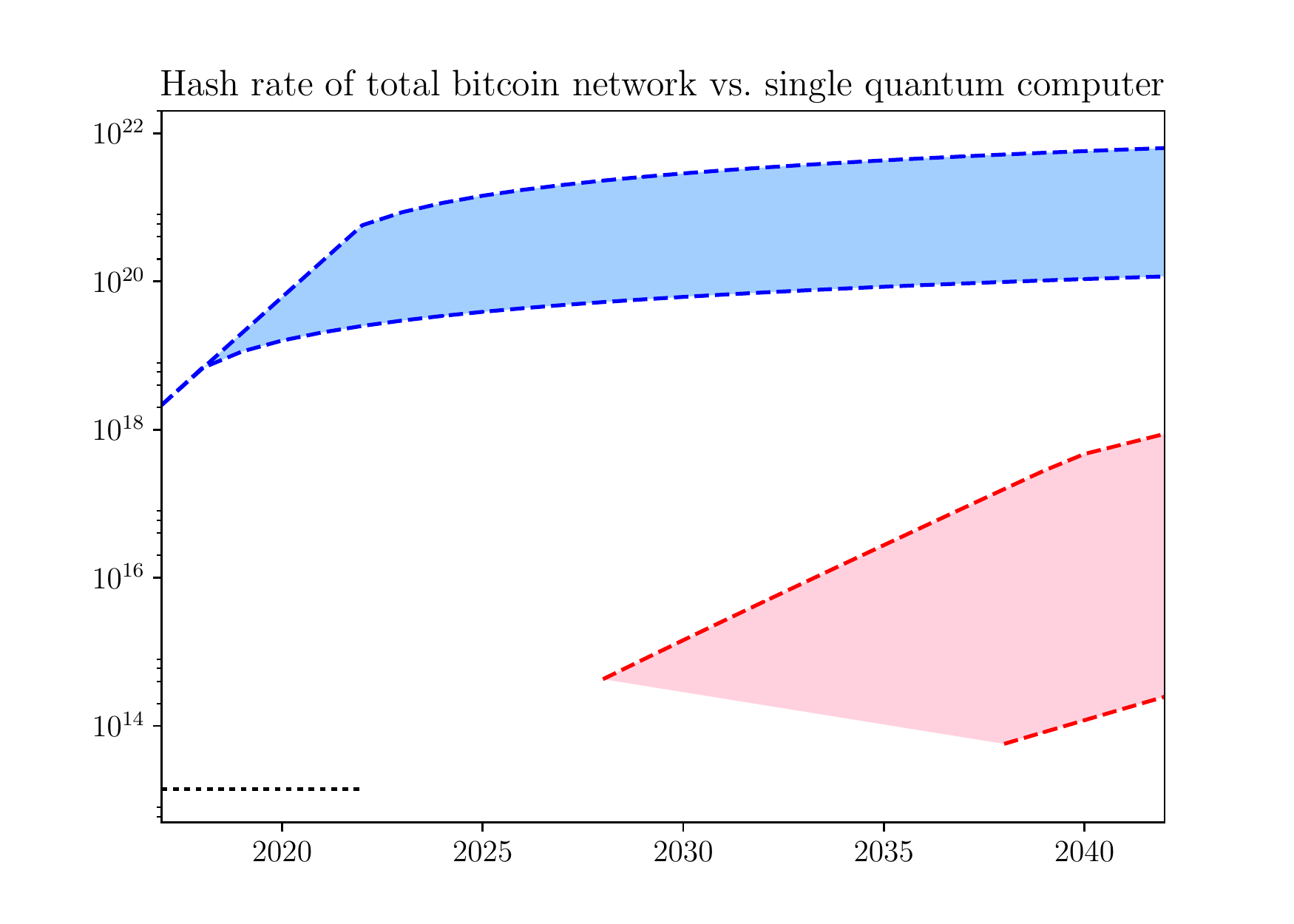}
  \caption{This plot shows two estimates of the hashing power (in hashes per second) of the bitcoin network (blue striped curves) vs.\ a single quantum computer (red striped curves) as a function of time for the next 25 years. We give more and less optimistic estimates and uncertainty regions (blue and orange area). The model is described in detail in Appendices~\ref{app:model1} and~\ref{app:model2}. Prior to 2028 (in the more optimistic estimate) there will not be any quantum computer with sufficiently many qubits to implement the Grover algorithm. For comparison, the black dotted line shows the hash rate of a single ASIC device today.}
  \label{fig:hash}
\end{figure}
The number of logical qubits needed in the Grover algorithm is fixed at $2402$, independent of the difficulty. The number of physical qubits needed is
\[
n_Q=2402\times c_{n_Q}(D,p_g),
\]
where $c_{n_Q}$ is the overhead in space, i.e. physical qubits, incurred due to quantum error correction, and is also a function of difficulty and gate error rate. 

In appendix \ref{app:overheads} we show how to calculate the overheads in time and space incurred by error correction. The results showing the performance of a quantum computer for blockchain attacks are given in Fig. \ref{fig:mineattack}. To connect these results to achievable specifications, we focus on superconducting circuits which as of now have the fastest quantum gate speeds amoung candidate quantum technologies and offer a promising path forward to scalability. Assuming maximum gate speeds attainable on current devices of $s=66.7$MHz \cite{Kirchhoff:17} and assuming an experimentally challenging, but not implausible, physical gate error rate of $p_g=5\times 10^{-4}$, and close to current difficulty $D=10^{12}$, the overheads are $c_{\tau}=538.6$ and $c_{n_Q}=1810.7$,  implying an effective hash rate of $h_{\rm QC}=13.8$GH/s using $n_Q=4.4\times 10^6$ physical qubits. This is more than one thousand times slower than off the shelf ASIC devices which achieve hash rates of $14$TH/s \footnote{Using e.g. the AntMiner S9 available at \href{https://asicminermarket.com/product/antminer-s9-14t-1600w-psu-14ths-2-fan-2/}{https://asicminermarket.com}}; the reason being the slow quantum gate speed and delays for fault tolerant $T$ gate construction.  
 
Quantum technologies are posed to scale up significantly in the next decades with a quantum version of Moore's law likely to take over for quantum clock speeds, gate fidelities, and qubit number. Guided by current improvements in superconducting quantum circuit technology, forecasts for such improvements are given in Appendices~\ref{app:model1} and~\ref{app:model2}. This allows us to estimate of the power of a quantum computer as a function of time as shown in Figure~\ref{fig:hash}. Evidently, it will be some time before quantum computers outcompete classical machines for this task and when they do, a single quantum computer will not have majority hashing power. However, even given modest improvement in power over competing miners' classical machines, certain attacks become more profitable such as attacks on mining pools that use smart contracts to reward pool miners to withhold their blocks \cite{Velner:17}. For example, given the optimistic assumptions outlined in Appendix \ref{app:overheads} where the effective hash rate scales like $h_{\rm QC}=0.04\times s \sqrt{D}$, at $D=10^{13}$ and $s=50$GHz, the fractional hashing power of a group of $20$ quantum machines running in parallel relative to total hashing power is $0.1\%$. This would allow for a quantum attack on pool mining with minimal bribing cost that would reduce pool revenue by $10\%$.

\subsection{Attacks on signatures}

Signatures in bitcoin are made using the Elliptic Curve Digital Signature Algorithm based on the secp256k1 curve. The security of this 
system is based on the hardness of the Elliptic Curve Discrete Log Problem (ECDLP).  While this problem is still believed to be hard 
classically, an efficient quantum algorithm to solve this problem was given by Shor~\cite{shor99}.  This algorithm means that a sufficiently 
large universal quantum computer can efficiently compute the private key associated with a given public key rendering this scheme 
completely insecure.  The implications for bitcoin are the following:
\begin{enumerate}
\item (Reusing addresses) To spend bitcoin from an address the public key associated with that address must be revealed.  Once the public 
key is revealed in the presence of a quantum computer the address is no longer safe and thus should never be used again.  While always using fresh addresses is already the suggested practice in Bitcoin, in practice this is not always followed.  Any address that has bitcoin and for which the public key 
has been revealed is completely insecure. 
\item (Processed transactions) If a transaction is made from an address which has not been spent from before, and this transaction is placed 
on the blockchain with 
several blocks following it, then this transaction is reasonably secure against quantum attacks.  The private key could be derived from the 
published public key, but as the address has already been spent this would have to be combined with out-hashing the network to 
perform a double spending attack.  As we have 
seen in Section~\ref{miningattack}, even with a quantum computer a double spending attack is unlikely once the transaction has many blocks 
following it.
\item (Unprocessed transactions) After a transaction has been broadcast to the network, but before it is placed on the blockchain it is at 
risk from a quantum attack.  If the secret key can be derived from the broadcast public key before the transaction is placed on the 
blockchain, then an attacker could use this secret key to broadcast a new transaction from the same address to his own address.  If the 
attacker then ensures that this new transaction is placed on the blockchain first, then he can effectively steal all 
the bitcoin behind the original address.  
\end{enumerate}

 \begin{figure}[t]
   \begin{center}
     \subfigure[]{
       \includegraphics[width=0.455\textwidth]{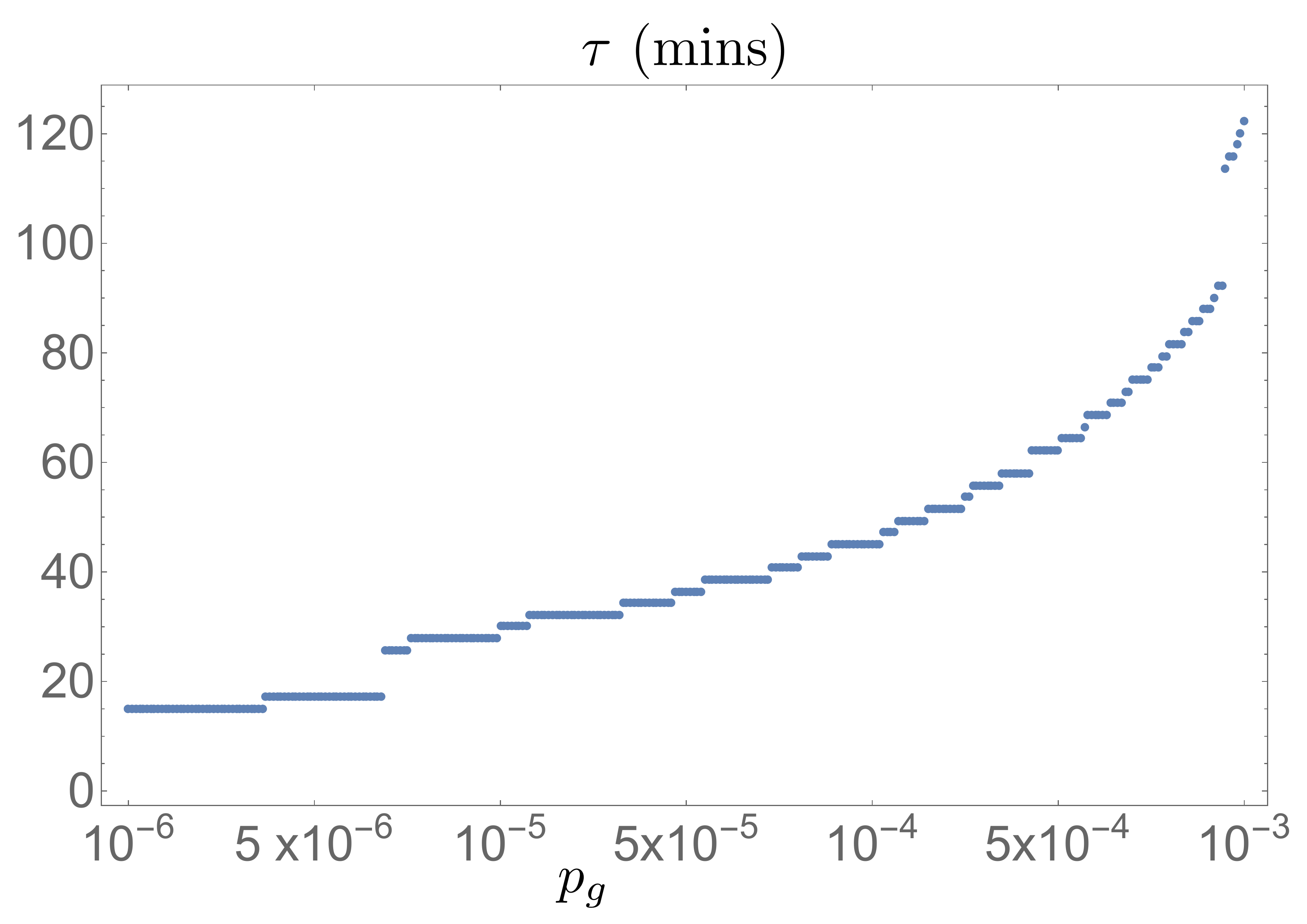}
       %\label{fig:numQ}
       }
     \subfigure[]{
       \includegraphics[width=0.476\textwidth]{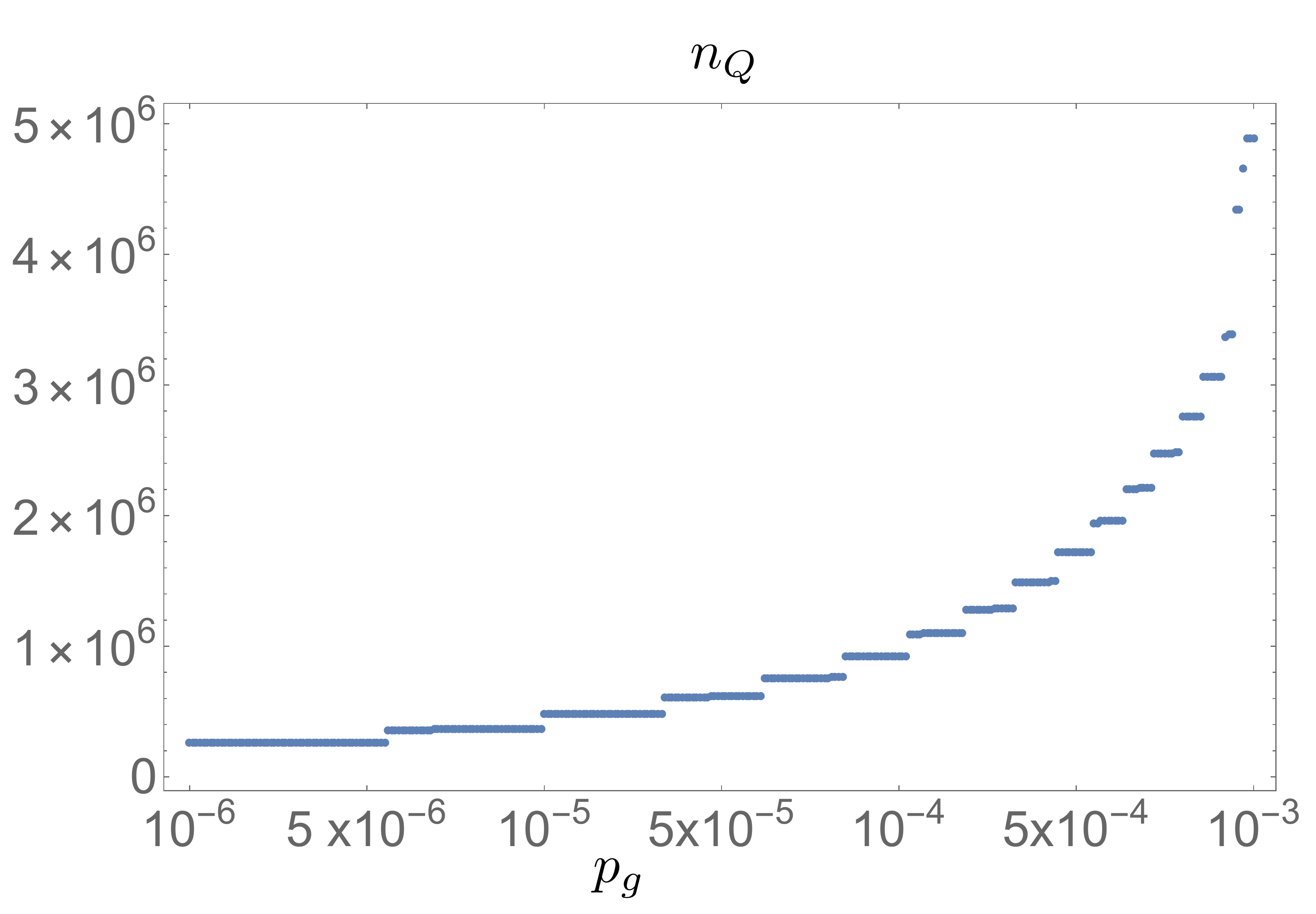}
       %\label{fig:hashrateQC}
       }
   \end{center}
   \caption{Performance of a quantum computer operating at $10$GHz clock speed for attacks on digital signatures using the elliptic curve digital signature algorithm. (a) Time in minutes to break the signature as a function of physical gate error rate $p_g$. (b) Number of physical qubits used by the quantum computer.}
   \label{fig:sigattack}
 \end{figure}
 
We view item~(3) as the most serious attack.  To determine the seriousness of this attack it is important to precisely estimate how much 
time it would take a quantum computer to compute the ECDLP, and if this could be done in a time close to the block interval.

For an instance with a $n$ bit prime field, a recently optimized analysis shows a quantum computer can solve the problem using $9n+2\lceil \log_2(n)\rceil+10$ logical qubits and $(448\log_2(n)+4090)n^3$ Toffoli gates \cite{Roetteler:17}. Bitcoin uses $n=256$ bit signatures so the number of Toffoli gates is $1.28\times 10^{11}$, which can be slightly parallelized to depth $1.16\times 10^{11}$. Each Toffoli can be realized using a small circuit of $T$ gate depth one acting on $7$ qubits in parallel (including $4$ ancilla qubits) \cite{Selinger:2013cr}. 

Following the analysis of Sec. \ref{miningattack}, we can estimate the resources needed for a quantum attack on the digital signatures. As with block mining, the dominant time is consumed by distilling magic states for the logical $T$ gates. 
 The time to solve the ECDLP on a quantum processor is
\[
\tau=1.28\times 10^{11}\times c_{\tau}(p_g)/s,
\]
where the time overhead $c_{\tau}$ now only depends on gate error rate, and $s$ is again the clock speed.
The number of physics qubits needed is
\[
n_Q=2334\times c_{n_Q}(p_g),
\]
where the first factor is the number of logical qubits including 4 logical ancilla qubits, and $c_{n_Q}$ is the space overhead.

The performance of a quantum computer to attack digital signatures is given in Fig. \ref{fig:sigattack}.
Using a surface code with a physical gate error rate of $p_g=5\times 10^{-4}$, the overhead factors are $c_{\tau}=291.7$ and $c_{n_Q}=735.3$, and the time to solve the problem at $66.6$ MHz clock speed  is $6.49$ days using $1.7\times 10^6$ physical qubits. Looking forward to performance improvements, for $10$GHz clock speed and error rate of $10^{-5}$, the signature is cracked in $30$ minutes using 485550 qubits. The latter makes the attack in item~(3) quite 
possible and would render the current Bitcoin system highly insecure. 
An estimate of the time required for a quantum computer to break the signature scheme as a function of time is given in Figure~\ref{fig:signature}, based on the model described in Appendices~\ref{app:model1} and~\ref{app:model2}.

\begin{figure}[t]
  \includegraphics[width=0.75\textwidth]{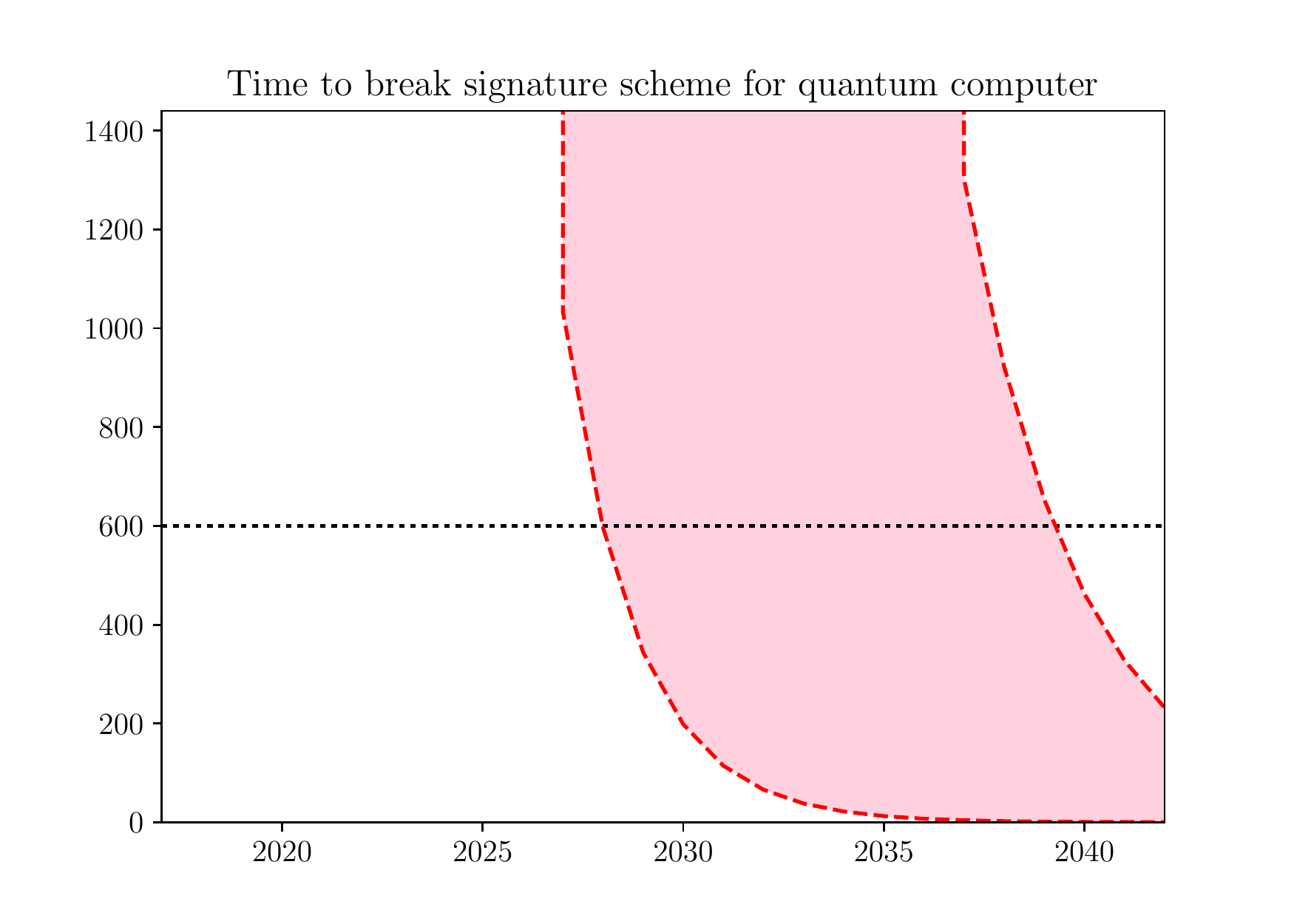}
  \caption{This plot shows two estimates of the time (in seconds) required for a quantum computer to break the signature scheme (red curves) as a function of time for the next 25 years. We give more and less optimistic estimates (red striped lines). The models are described in detail in Appendix~\ref{app:model2}. According to this estimate, the signature scheme can be broken in less than 10 minutes (600 seconds, black dotted line) as early as 2027.}
  \label{fig:signature}
\end{figure}

\subsection{Future enhancements of quantum attacks}
We have described attacks on the Bitcoin protocol using known quantum algorithms and error correction schemes. While some of the estimates for quantum computing speed and scaling may appear optimistic, it is important to keep in mind that there are several avenues for improved performance of quantum computers to solve the aforementioned problems. 

First, the assumed error correction code here is the surface code which needs significant classical computational overhead for state distillation, error syndrome extraction, and correction. Other codes which afford transversal Clifford \emph{and} non-Clifford gates could overcome the need for slow state distillation \cite{Paetznick:2013le}. In fact the slow down from classical processing for syndrome extraction and correction could be removed entirely using a measurement free protocol, see e.g. \cite{Paz-Silva:2010ek} which in a recent analysis show error thresholds  \cite{Crow:2016it} only about $5$ times worse than the measurement heavy surface code. This could potentially dramatically improve overall speed of error correction. 

Second, reductions in logical gate counts of the quantum circuits are possible as more efficient advanced quantum-computation techniques are being developed. For example, using a particular large-size example 
problem (including oracle implementations) that was analyzed in a previous work \cite{Scherer:2017fk}, a direct comparison of the concrete gate counts,
obtained by the software package Quipper, has been achieved between the old \cite{Harrow:2009yq} and the new \cite{Childs:15} linear-systems solving quantum 
algorithms, showing an improvement of several orders of magnitude \footnote{A. Scherer, personal communication}. Given that the quantum Shor and Grover algorithms have been well studied and highly optimized, one would not expect such a dramatic improvement, nonetheless it is likely some improvement is possible.   

Third, different quantum algorithms might provide relative speedups. Recent work by Kaliski \cite{Kaliski:17}, presents a quantum algorithm for the Discrete Logarithm Problem: find $m$ given $b=a^m$, where $b$ is a known target value and $a$ is a known base, using queries to a so called ``magic box" subroutine which computes the most significant bit of $m$. By repeating queries using judiciously chosen powers of the target value, all bits of $m$ can be calculated and the problem solved. Because different bits are solved one by one the problem can be  distributed to multiple quantum processors. Each processor requires a number of logical qubits comparable to solving the entire problem, but the overall time would be reduced by the parallelization. Furthermore, the overhead for quantum error correction is likely reduced as the phases in the quantum fourier transform part of the circuit need not be as accurate as in the original Shor algorithm.   

\section{Countermeasures}
\label{sec:counter}

\subsection{Alternative proofs-of-work}
As we have seen in the last section, a quantum computer can use Grover search to perform the bitcoin proof-of-work 
using quadratically fewer hashes than are needed classically.  In this section we investigate alternative proofs-of-work that 
might offer less of a quantum advantage.  The basic properties we want from a proof-of-work are:
\begin{enumerate}
\item (Difficulty) The difficulty of the problem can be adjusted in accordance with the computing power available in the network.
\item (Asymmetry) It is much easier to verify the proof-of-work has been successfully completed than to perform the proof-of-work.
\item (No quantum advantage) The proof-of-work cannot be accomplished significantly faster with a quantum computer than with a classical 
computer.
\end{enumerate}
The bitcoin proof-of-work accomplishes items~(1),(2), but we would like to find an alternative proof of work that does 
better on~(3).

Similar considerations have been investigated by authors trying to find a proof-of-work that, instead of~(3) look for 
proofs-of-work that cannot be accelerated by ASICs.  An approach to doing this is by looking at memory intensive proofs of work.  
Several interesting candidates have been suggested for this such as Momentum \cite{Larimer14}, based on finding collisions in a hash 
function, Cuckoo Cycle \cite{Tromp14}, based on finding constant sized subgraphs in a random graph, and 
Equihash \cite{BK17}, based on the generalized birthday problem.  These are also good candidates for a more quantum resistant 
proof-of-work.

These schemes all build on the hashcash-style proof-of-work and use the following template.  Let $h_1: \{0,1\}^* \rightarrow \{0,1\}^n$ 
be a cryptographically secure hash function and $H = h_1(\text{header})$ be the hash of the block header.  The goal is then to find 
a nonce $x$ such that 
\[
h_1(H \parallel x) \le t \text{ and } P(H,x) \enspace,
\]
for some predicate $P$.  The fact that the header and nonce have to satisfy the predicate $P$ means that the best algorithm 
will no longer simply iterate through nonces $x$ in succession.  Having a proof-of-work of this form also ensures that the parameter $t$ 
can still be chosen to vary the difficulty.

In what follows, we will analyse this template for the Momentum proof-of-work, as this can be related to known quantum lower bounds.
For the momentum proof of work, let $h_2: \{0,1\}^* \rightarrow \{0,1\}^\ell$ be another hash function with $n \leq \ell$.
In the original Momentum proposal $h_1$ can be taken as SHA-256 and $h_2$ as a memory intensive hash function, but this is less 
important for our discussion. The proof-of-work is to find $H, a,b$ such that 
\begin{equation}
\label{eq:momentum}
h_1(H \parallel a \parallel b) \le t \text{ and } h_2(H \parallel a) = h_2(H \parallel b) \text{ and } a,b \leq 2^{\ell} \enspace.
\end{equation}

First let's investigate the running time in order to solve this proof-of-work, assuming that the hash functions $h_1,h_2$ can be 
evaluated in unit time.  Taking a subset 
$S \subset \{0,1\}^{\ell}$ and evaluating $h_2(H \parallel a)$ for all $a \in S$, we expect to find about $|S|^2/2^{\ell}$ many 
collisions.  Notice that by using an appropriate data structure, these collisions can be found in time  
about $|S|$.

One algorithm is then as follows.  For each $H$, we evaluate $h_2$ on a subset $S$ and find about $|S|^2/2^{\ell}$ 
many pairs 
$a,b$ such that $h_2(H \parallel a) = h_2(H \parallel b)$.  For each collision we then test 
$h_1(H \parallel a \parallel b) \le t$.  In expectation, we will have to perform this second test $2^n/t$ many times.  Thus 
the number of $H$'s we will have to try is about $m = \max \{1, \tfrac{2^{n+\ell}}{t|S|^2}\}$, since we have to try at least one $H$.
As for each $H$ we spend time 
$|S| $, the total running time is 
$m|S| $. We see that it is the smallest when $|S| =  \sqrt{\tfrac{2^{n+\ell}}{t}}$, that is when $m = 1$,
and we just try one $H$.
This optimal running time is then $T = \sqrt{\tfrac{2^{n+\ell  }}{t}} $, and to achieve it we have to use a memory 
of equal size to the running time,
which might be prohibitive. For some smaller memory $|S| <  \sqrt{\tfrac{2^{n+\ell}}{t}}$ the running time will be 
$ \tfrac{2^{n+\ell+1}}{t|S|}$.

Now let us look at the running time on a quantum computer.  
On a quantum computer we can do the following.  Call $H$ 
\emph{good} if there exists $a,b \in S$ such that 
$h_1(H \parallel a \parallel b) \le t \text{ and } h_2(H \parallel a) = h_2(H \parallel b)$.  
Testing if an $H$ is good requires finding a collision, and therefore necessitates at least $|S|^{2/3}$ time 
by the quantum query lower bound of Aaronson and Shi \cite{AS04}.  Note that this lower bound is tight as 
finding such a collision can also be done in roughly $|S|^{2/3}$ time using Ambainis' element distinctness algorithm \cite{Amb07}.
We have argued above that a set of size $m = \max \{1, \tfrac{2^{n+\ell}}{t|S|}\}$ 
is needed 
to find at least one good $H$. By the optimality of Grover search \cite{BBBV97} we know that we have to perform at least 
$\sqrt{m}$ many tests to find a good $H$.  As testing if an $H$ is good requires time $|S|^{2/3}$, the total running time is at least
$\sqrt{m} |S|^{2/3}$.  
As the classical running time is $m|S|$, 
we see that unlike for the current proof of work in Bitcoin, 
with this proposal a quantum computer would not be able to achieve a quadratic advantage as soon as 
$S$ is more than constant size. In particular, 
since $\sqrt{m} |S|^{2/3}$ is minimized also when $S =  \sqrt{\tfrac{2^{n+\ell}}{t}}$, the running time of
even the fastest  quantum algorithm
is at least $T^{2/3}$, which is substantially larger than $T^{1/2}$.

\subsection{Review of post-quantum signature schemes}

Many presumably quantum-safe public-key signature schemes have been proposed in the literature.  Some examples of these are hash-based signature schemes (LMS~\cite{LM95}, XMSS~\cite{BDH11}, SPHINCS~\cite{BHHLNPSSW15},NSW~\cite{NSW05}), code-based schemes (CFS~\cite{CFS01}, QUARTZ~\cite{PCG01}), schemes based on multivariate polynomials (RAINBOW~\cite{DS05}), and lattice-based schemes (GPV~\cite{GPV08}, LYU~\cite{LYU12}, BLISS~\cite{LDLL13}, DILITHIUM~\cite{DLLSSS17}, NTRU~\cite{MBDG14}). Each of these cryptosystems have varying degree of efficiency. For a comparison in terms of signature size and key size, 
see Table \ref{sig_table}.

In the blockchain context the most important parameters of a signature scheme are the signature and public key lengths, as these must 
be stored in some capacity to fully verify transactions, and the time to verify the signature.   Looking at Table~\ref{sig_table}, with respect 
the sum of signature and public key lengths, the only reasonable options are hash and lattice based schemes.  

Hash based schemes like XMSS have the advantage of having provable security, at least assuming the chosen hash function behaves 
like a random oracle.  The generic quantum attack against these schemes is to use Grover's algorithm which means that their quantum 
security level is half of the classical security level.  In contrast, the best known quantum attack against DILITHIUM at 138 bit classical 
security level requires time $2^{125}$.  Thus at the same level of \emph{quantum} security, lattice based schemes have some advantage in 
signature plus public key length.

Although the lattice based scheme BLISS has the shortest sum of signature and public key lengths of all the schemes in 
Table~\ref{sig_table}, there are some reasons not to choose BLISS in practice.  The
security of BLISS relies on hardness of the NTRU 
problem and the assumption that solving this problem is equivalent to finding a short vector in a so-called NTRU lattice. It has been shown recently~\cite{KirFou17} that this assumption might be too optimistic, at least 
for large parameters. Moreover, there is a history of attacks on prior NTRU-based signature schemes~\cite{NR06,DN12}.  Perhaps most 
fatally, BLISS is difficult to implement in a secure way as it is very susceptible to side channel attacks.  The production grade strongSwan 
implementation of BLISS has been attacked in this way by Pessl et al. \cite{PBY17}, who showed that the signing key could be recovered 
after observing about 6000 signature generations.  

\begin{table}
\centering
\begin{tabular}{|c|c|c|c|c|c|}
\hline
Type & Name & security level (bits) & PK length (kb) & Sig. length (kb) & PK + Sig. lengths (kb) \\
\hline
I.1  & GPV  & 100 & 300 & 240  & 540 \\
I.2  & LYU   &100  & 65 & 103 & 168 \\
I.3 & BLISS & 128 & 7 & 5 & 12 \\
I.4 & DILITHIUM & 138 & 11.8 & 21.6 & 33.4 \\
II.1 & RAINBOW & 160 & 305 & 0.244 & 305 \\
III.1 & LMS  & 128  & 0.448  & 20 & 20.5 \\
III.2 & XMSS & 128 & 0.544 & 20 & 20.5 \\
III.3 & SPHINCS & 128 & 8 & 328 & 336 \\
III.4 & NSW & 128 & 0.256 & 36 & 36 \\
IV.1 & CFS & 83 & 9216 & 0.1 & 9216 \\
IV.2 & QUARTZ  & 80 & 568 & 0.128 & 568 \\
\hline
\end{tabular}
\caption{Comparison of the public key (PK) and signature lengths of post-quantum signature schemes in kilobits (kb).  The security 
level given is against classical attacks.  Type~I are lattice based, type~II based on multivariate polynomials, type~III hashing based, 
and type~IV code based.}
\label{sig_table}
\end{table}

\appendix

\section{Estimating error correction resource overheads for quantum attacks}
\label{app:overheads}

Here we describe how the overhead factors for quantum error correction are calculated in order to obtain resource costs for quantum attacks on blockchains and digital signatures.
The method follows the analysis given in Refs. \cite{PhysRevA.86.032324, AMGMPS:2016}. We first determine $n_T$ and $n_C$, the number of $T$ gates and Clifford gates respectively needed in the algorithm. The pseudo-code to compute the overhead is given in Table \ref{tab:resourcecalc}.
For the blockchain attack on $n_L=2402$ qubits, these values are
\[
n_T=297784\times \pi 2^{14}\sqrt{10\cdot D},\quad n_C=29.4\times n_T.
\]
For the Digital Signature attack on $n_L=2334$ qubits, the values are \footnote{The factor of $20$ for the number of Clifford gates per $T$ gate is based is based on the construction of $T$ gate depth one representations of the Toffoli gate in \cite{Selinger:2013cr}.}
\[
n_T=1.28\times 10^{11},\quad n_C=20\times n_T .
\]

\begin{table}
\begin{algorithmic}
\Function{CalculateFactoryResources}{$p_g$, $n_T$}
\Comment{iterates layers of error correction in factory}
\State $p_{\textrm{tol}} \gets \frac{1}{n_T}$
\Comment{(uncorrected) error tolerance}
\State $i \gets 0$
\While{$p_{\textrm{tol}} <10p_g $}
  \State $i \gets i + 1$
  \Comment{add layer}
  \State $d_{i} \gets  \min \Big\{ d \in \mathbb{N} : 192d \cdot (100 p_g)^{\frac{d+1}{2}} 
  \geq \frac{p_{\textrm{tol}}}{2} \Big\}$
  \Comment{code distance in this layer}
  \State $p_{\textrm{tol}} \gets (\frac{p_{\textrm{tol}}}{70})^{\frac13}$
  \Comment{increased error tolerance}
\EndWhile
\State layers $\gets i$
\State $\tau \gets n_T \cdot 10 \sum_{i=1}^{\textrm{layers}} d_i$
\Comment{total clock cycles (only counts $T$ gates)}
\State $Q_{\textrm{factory}} \gets 50 (d_\textrm{layers})^2 \cdot 15^{\textrm{layers}-1}$
\Comment{total physical qubits for factory}
\State \Return $(\tau, Q_{\textrm{factory}})$
\EndFunction
\end{algorithmic}  
 
\begin{algorithmic}
\Function{CalculateCircuitResources}{$p_g$, $n_C$, $n_L$} 
\State $d_C \gets \min \Big\{ d \in \mathbb{N} : (80 p_g)^{\frac{d+1}{2}} \geq \frac{1}{n_C} \Big\}$
\Comment{code distance for circuit (single layer)}
\State \Return $Q_{\textrm{circuit}} \gets 3.125 n_L d_C$
\Comment{total physical qubits for circuit}
\EndFunction
\end{algorithmic}   
\caption{\label{tab:resourcecalc}
Algorithms to compute space and time resources for quantum attacks. The inputs are $p_g$, the physical gate error rate; $n_C$, the total number of Clifford gates in the logical circuit; $n_T$, the total number of $T$ gates in the logical circuit; and $n_L$, the number of logical qubits. The outputs are $\tau$, the time cost in number of clock cycles; and $n_Q=Q_{\textrm{circuit}} + Q_{\textrm{factory}}$, the number of physical qubits used for the computation including state distillation.}
\end{table}

If we look some years into the future we can speculate as to plausible improvements in quantum computer technology. If we assume a quantum error correction code that supports transversal Clifford and non-Clifford gates so there is no distillation slow down and that it is done in a measurement free manner so that no classical error syndrome processing is necessary, then the number of cycles needed for one oracle call is determined solely by the circuit depth which is $2142094$. This is based on an overall circuit depth calculated as follows. The oracle invokes two calls to the SHA256 hash function, and this is done twice, once to compute it and again to uncompute it. Each hash has a reversible circuit depth of $528768$. Similarly, there are two multi-controlled phase gates used, one for inversion about the mean and one for the function call, each having a circuit depth $13511$, for a total depth $4\times 528768+2\times 13511=2142094$ (these numbers are from \cite{Suchara:EECS-2013-119} but could be further optimized). Then accepting potential overhead in space and physical qubit number, but assuming no time penalty for error correction or non Clifford gate distillation, this implies an improved effective hashing rate of 
 \[
 h_{\rm QC}=0.04\times s \sqrt{D}.
 \]
which is substantially faster.
For superconducting circuits, ultrafast geometric phase gates are possible at $\sim 50$ GHz, essentially limited by the microwave resonator frequency \cite{Romero:2012jy}. Using the above very optimistic assumptions, at difficulty $D=10^{12}$ the effective hash rate would be $h_{\rm QC}=2.0\times 10^3$TH/s.

\section{Modeling the development of bitcoin network hash rates and difficulty}
\label{app:model1}

\begin{figure}[t]
  \begin{center}
    \subfigure[]{
      \includegraphics[width=0.48\textwidth]{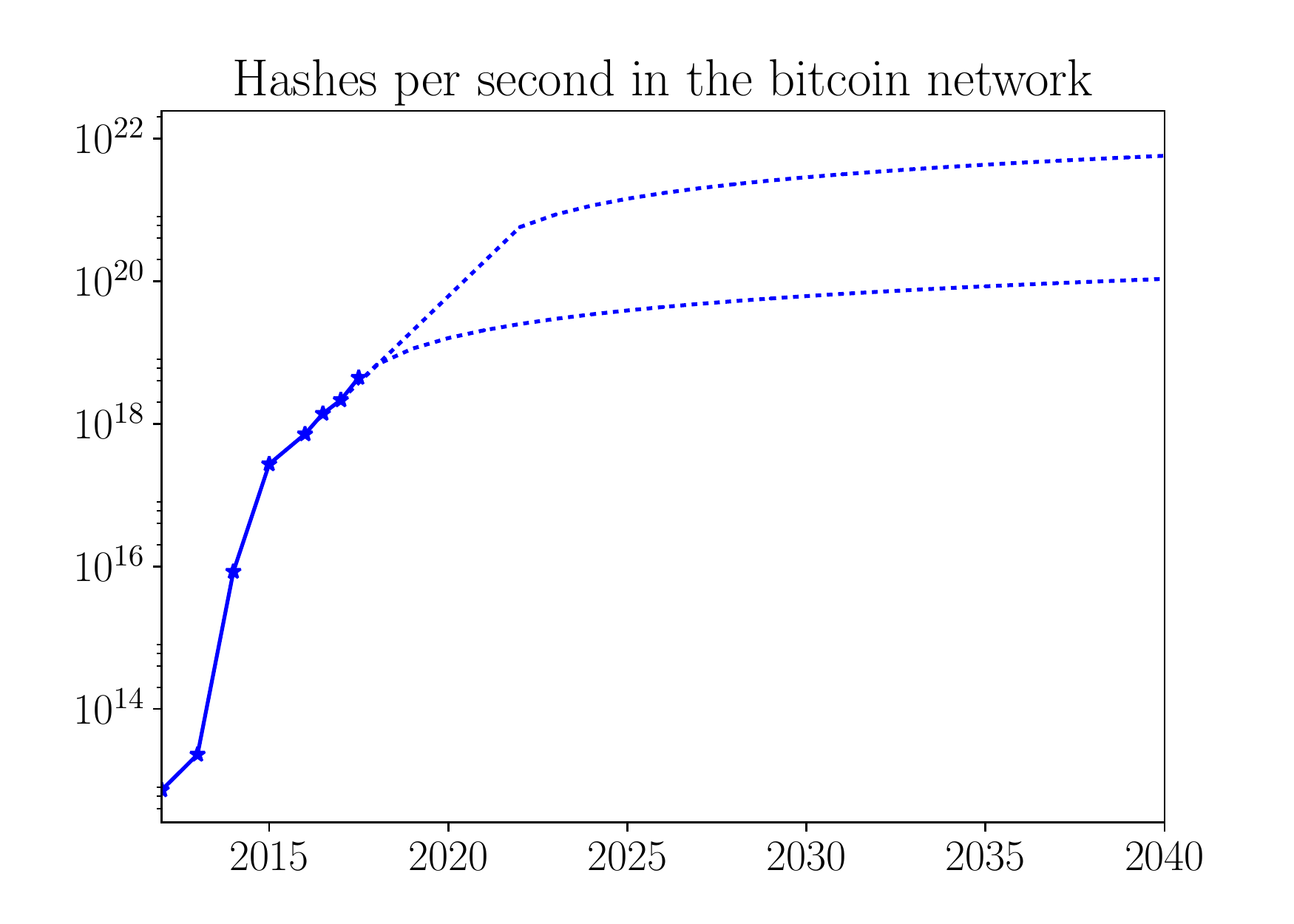}
      \label{fig:classump1}
    }
    \subfigure[]{
      \includegraphics[width=0.48\textwidth]{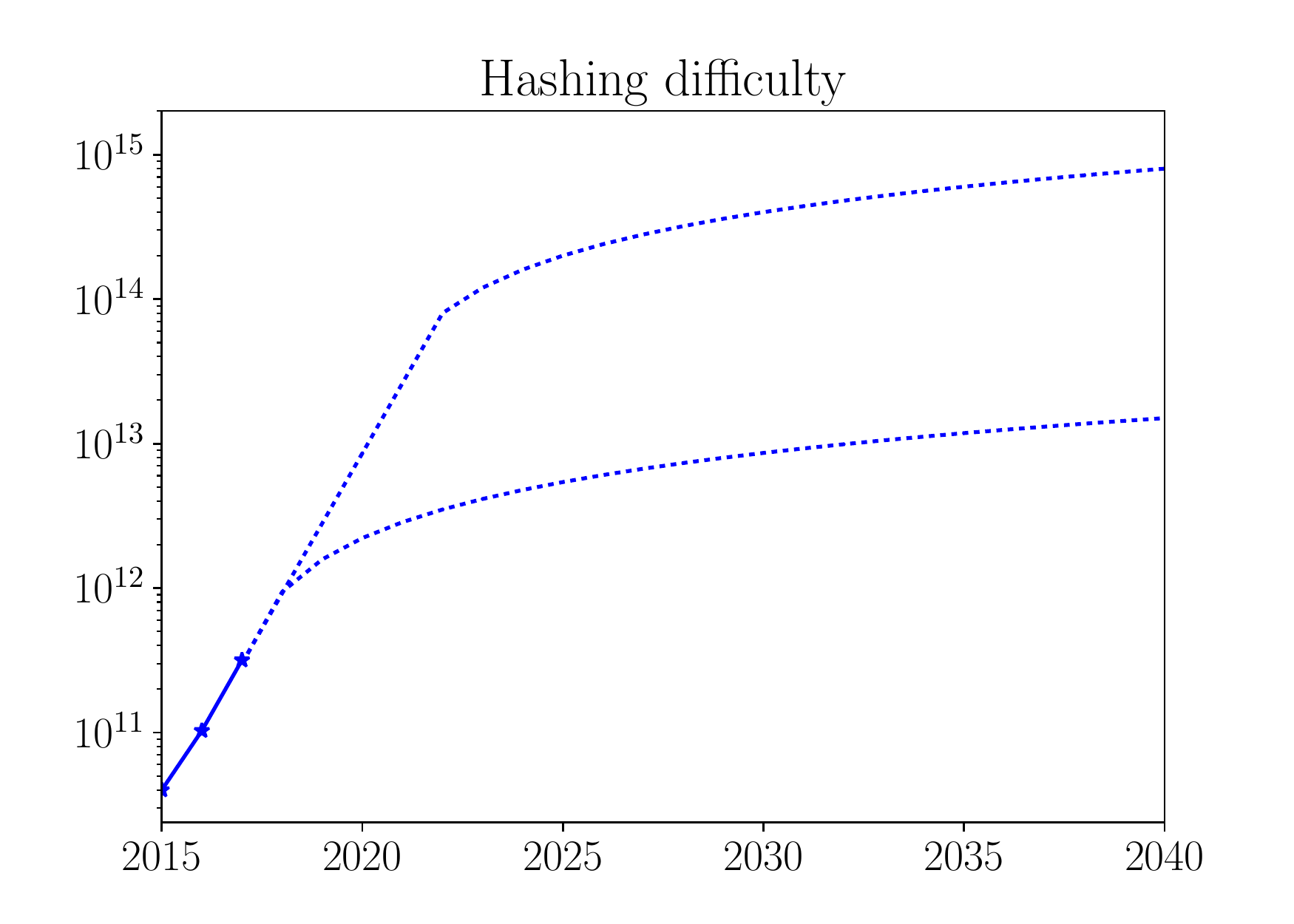}
      \label{fig:classump2}
      }
  \end{center}
  \caption{Prediction of the hash rate of the bitcoin network (in number of hashes per second) and the hashing difficulty as a function of time.}
  \label{fig:classump}
\end{figure}

The total number of hashes per second in the whole bitcoin network are taken from
\href{https://blockchain.info/charts/hash-rate?timespan=all&daysAverageString=7&scale=1}{blockchain.info}. The data points in Figure~\ref{fig:classump1} are the hash rates for the first of January (2012--2015) and first of January and July (2016--2017). The two dotted curves correspond to optimistic and less optimistic assumptions for the extrapolations. The optimistic extrapolation assumes that the present growth continues exponentially for five years and then saturates into a linear growth as the market gets saturated with fully optimized ASIC bitcoin miners. The less optimistic assumption assume linear growth at the present rate.

From the extrapolation of the bitcoin network hashrate we can determine the difficulty as a function of time. The expected number of hashes required to find a block in 10 minutes (600 seconds) is given by $\textrm{rate}(t) \cdot 600$, where $\textrm{rate}(t)$ is the total hash rate displayed in Figure~\ref{fig:classump1}. Thus the bitcoin hashing difficulty is calculated as $D(t) = \textrm{rate}(t) \cdot 600 \cdot 2^{-32}$ for the two scenarios discussed above. In Figure~\ref{fig:classump2} we compare this with values from \href{https://blockchain.info/charts/difficulty?timespan=all&scale=1}{blockchain.info} for the first of January of 2015--2017. 

\section{Modeling the development of quantum computers}
\label{app:model2}

There are several aspects of the development of quantum technologies that we must model. Since only few data points are available at this early stage of the development there is necessarily a lot of uncertainty in our estimates. We therefore give two different estimates, one that is optimistic about the pace of the development and another one that is considerably more pessimistic. Nonetheless, these predictions should be considered as a very rough estimate and might need to be adapted in the future.

\begin{figure}[t]
  \begin{center}
    \subfigure[]{
      \includegraphics[width=0.48\textwidth]{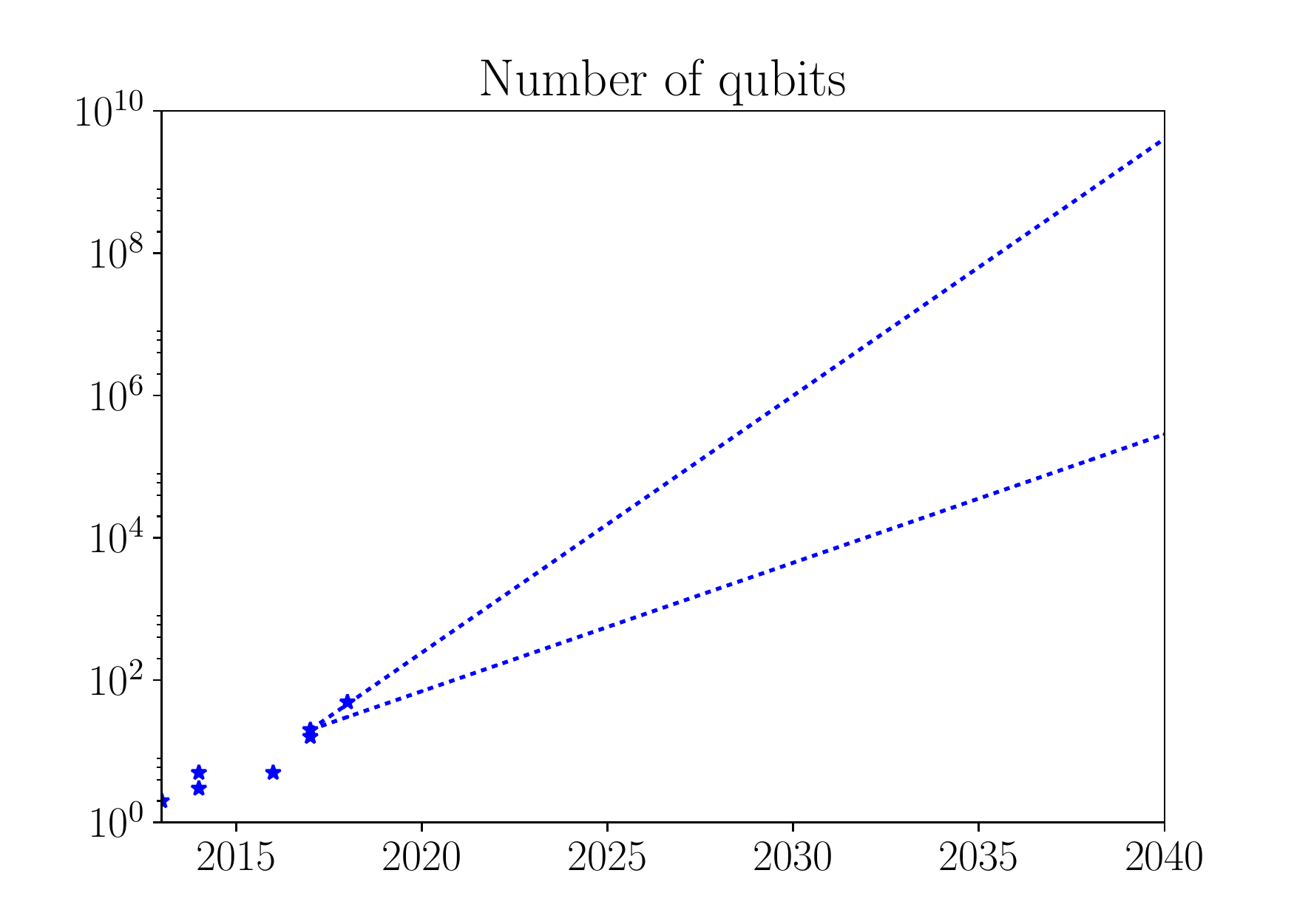}
      \label{fig:qassump1}
      }
    \subfigure[]{
      \includegraphics[width=0.48\textwidth]{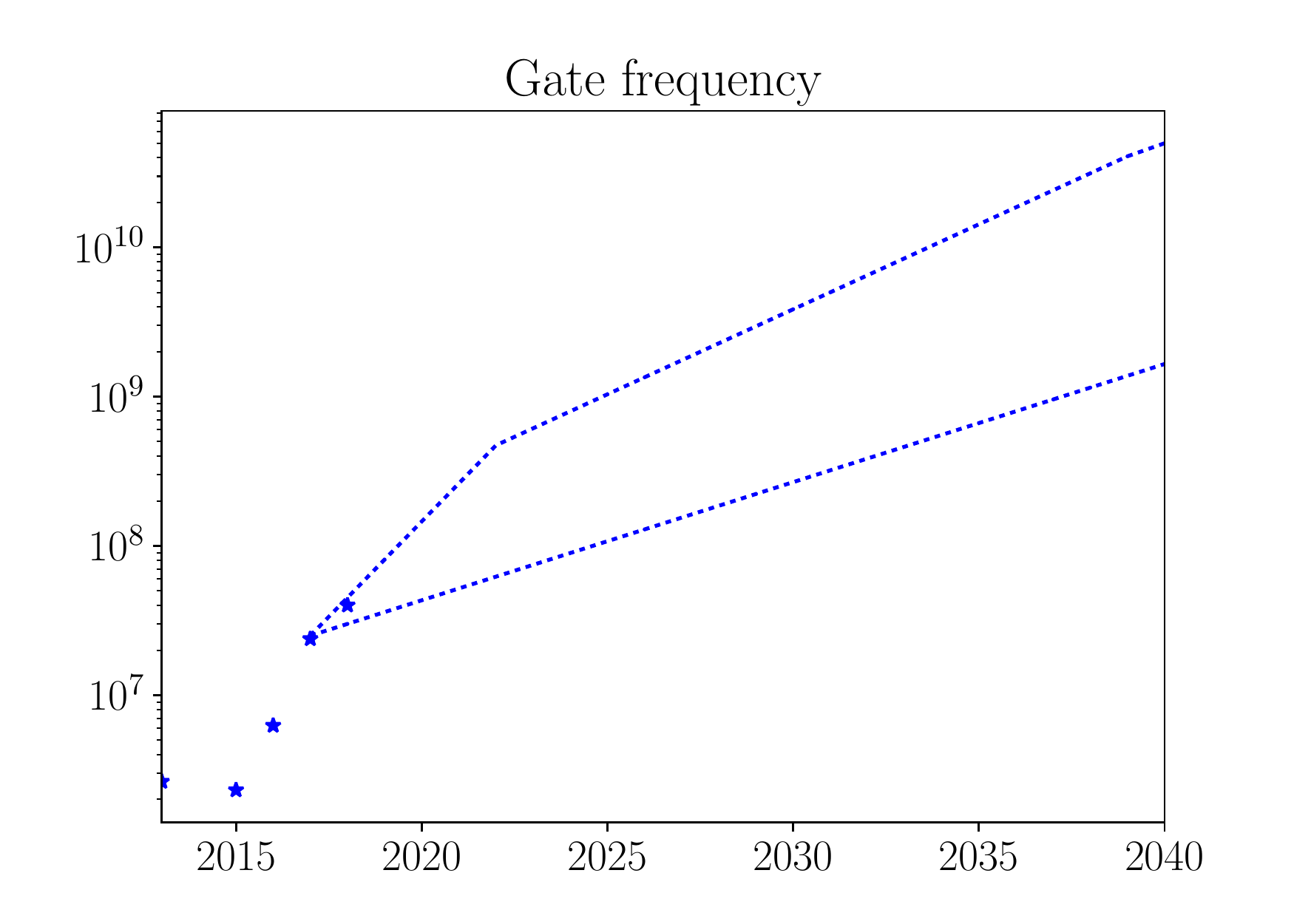}
      \label{fig:qassump2}
      }
    \subfigure[]{
      \includegraphics[width=0.48\textwidth]{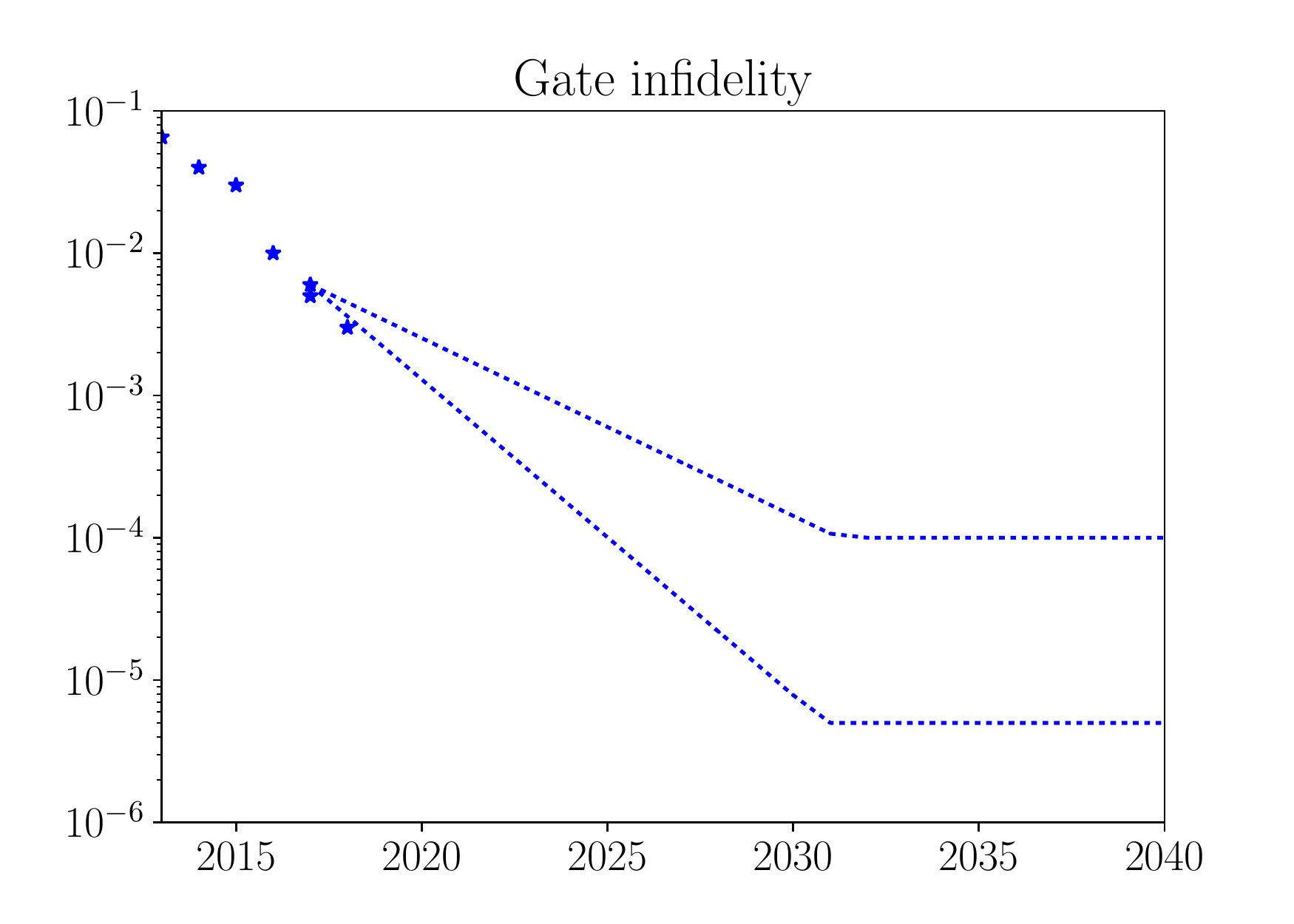}
      \label{fig:qassump3}
      }
    \subfigure[]{
      \includegraphics[width=0.48\textwidth]{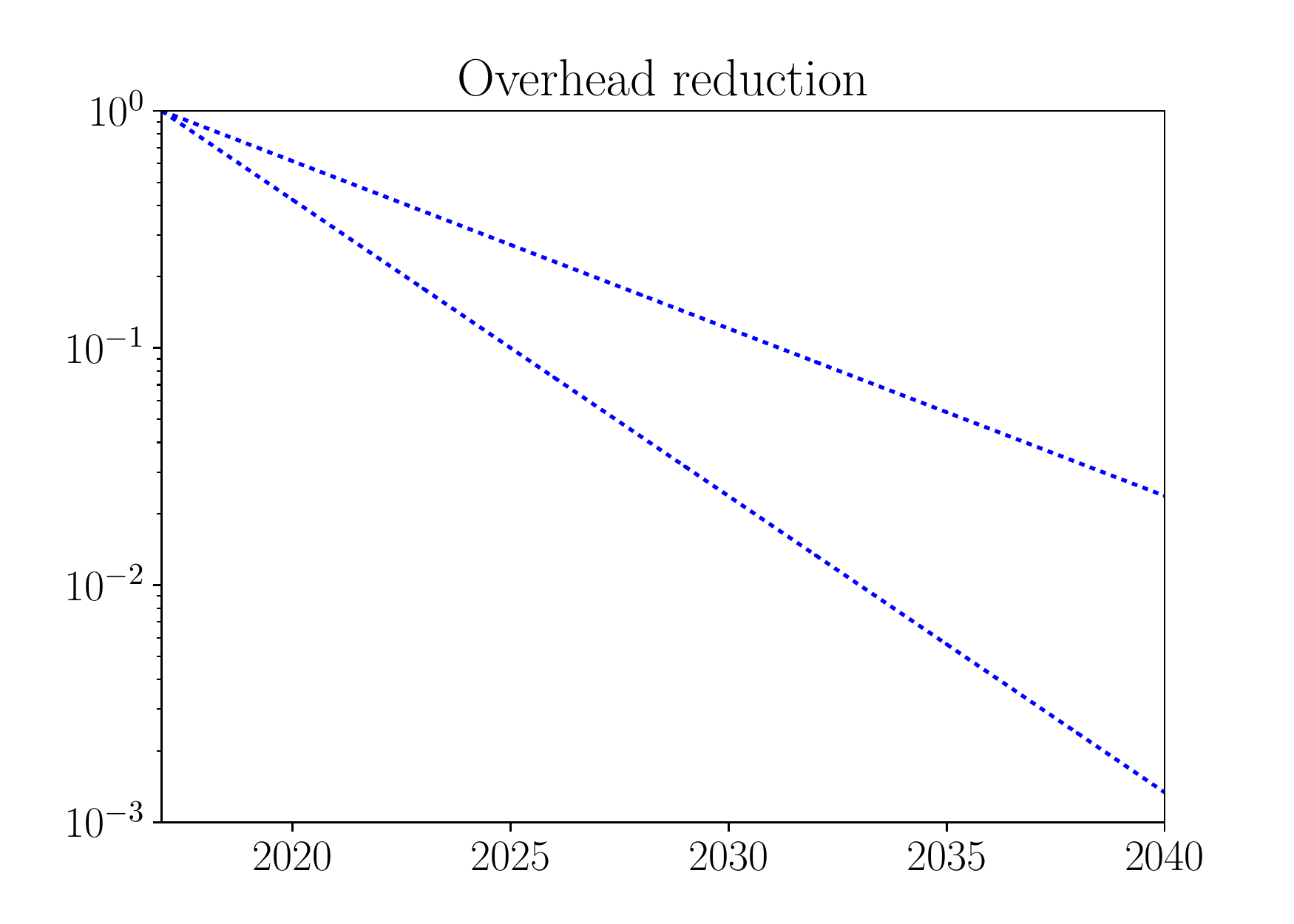}
      \label{fig:qassump4}
      }
  \end{center}
  \caption{Prediction of the number of qubits, the quantum gate frequency (in gate operations per second) and the quantum gate infidelity as a function of time. The forth plot models a reduction of the overhead due to theoretical advances.}
  \label{fig:qassump}
\end{figure}

First, we need to make an assumption on the number of qubits available at any point of time. As we focus only on solid state superconducting implementations there are only a few data points available. We assume that the number of available qubits will grow exponentially in time in the near future. The optimistic assumption is that the number will double every 10 months whereas the less optimistic assumption assumes the number doubles every 20 months. These two extrapolations are plotted in Figure~\ref{fig:qassump1}. The data points are taken from the following table:

\begin{center}
  \def\arraystretch{1.1}
  \setlength\tabcolsep{0.2cm}
  \begin{tabular}{r|l|l}
      number of qubits & year & reference \\
      \hline
      2  & 2013 & \cite{Corcoles:2013eq} \\
      5  & 2014 & \cite{barends14} \\
      3  & 2014 & \cite{chow14} \\
      5  & 2016 & IBM \\
      16 & 2017 & IBM \\
      20 & 2017 & \cite{Google2017} \\
      49 & 2018 & \cite{Google2017}
  \end{tabular}
\end{center}

We predict that the quantum gate frequency grows exponentially for the next years. This assumes that the classical control circuits will be sufficiently fast to control quantum gates at this frequencies. After a couple of years the growth slows down considerably because faster classical control circuits are necessary to further accelerate the quantum gates. We cap the quantum gate frequency at $50$ GHz (for the optimistic case) or $5$ GHz (for the less optimistic case), respectively, mostly because we expect that classical control circuits will not be able to control the quantum gates at higher frequencies. (See, e.g., \cite{herr11} for progress in this direction.)
This is shown in Figure~\ref{fig:qassump2}. The data points are taken from the following table:

\begin{center}
  \def\arraystretch{1.1}
  \setlength\tabcolsep{0.2cm}
  \begin{tabular}{r|l|l}
      gate time & year & reference \\
      \hline
      420ns & 2013 &  \cite{Corcoles:2013eq} \\
      433ns & 2015 & \cite{corcoles15} \\
      160ns & 2016 & \cite{sheldon16} \\
      42ns  & 2017 & \cite{Deng:2017fj} \\
      25ns  & 2018 & Google, projected for end of 2017
  \end{tabular}
\end{center}

The predicted development of the gate infidelity is shown in Figure~\ref{fig:qassump3}. We assume that the gate infidelity will continue to drop exponentially but that this development will stall at an infidelity of $5 \cdot 10^{-6}$ (optimistic case) or $5 \cdot 10^{-5}$ (less optimistic case). For the optimistic case we expect that the gate infidelity will continue to follow DeVincenzo's law which predicts a reduction of the infidility by a factor of 2 per year. The data points are taken from the following table:

\begin{center}
  \def\arraystretch{1.1}
  \setlength\tabcolsep{0.2cm}
  \begin{tabular}{r|l|l}
      gate fidelity & year & reference \\
      \hline
      0.9347 & 2013 & \cite{Corcoles:2013eq} \\
      0.96   & 2014 & \cite{chow14} \\
      0.97   & 2015 & \cite{corcoles15} \\
      0.99   & 2016 & \cite{sheldon16} \\
      0.995  & 2017 & \cite{Google2017} \\ 
      0.997  & 2018 & \cite{Google2017}
   \end{tabular}
\end{center}

Finally, we assume that the number of qubits and time steps required by any algorithm will be reduced over time for two reasons. First, the gate fidelity will increase over time and thus allow for more efficient fault-tolerant schemes to be used. Second, theoretical advances will allow to decrease the number of qubits and gates required to implement the algorithm and fault-tolerant schemes. We expect that this factor will be $\textrm{overhead}(t) = \beta^{t - 2017}$ where $\beta \in \{0.75, 0.85\}$ for optimistic and less optimistic assumptions, respectively.

 \section*{Acknowledgements}
 MT and GB would like to thank Michael Bremner for initial discussions. TL would like to thank John Tromp for helpful comments and discussions about proof-of-work and Ronald de Wolf for conversations about
 parallel quantum search.

\bibliographystyle{alpha}
\bibliography{library_GKB.bib,library_mt.bib,library_da.bib,blockchain.bib}

\end{document}